# TOWARDS COMMISSIONING THE FERMILAB MUON G-2 EXPERIMENT*

D. Stratakis[†], M. E. Convery, J. P. Morgan, M. J. Syphers[1] Fermi National Accelerator Laboratory, Batavia IL, USA M.
M. Korostelev, University of Lancaster/Cockcroft Institute, UK
A. Fiedler, Northern Illinois University, DeKalb IL, USA
S. Kim, Cornell University, Ithaca NY, USA
J. D. Crnkovic, W. M. Morse, Brookhaven National Laboratory, Upton NY, USA
[1]also at Northern Illinois University, DeKalb IL, USA

*Abstract*

Starting this summer, Fermilab will host a key experiment dedicated to the search for signals of new physics: The Fermilab Muon g-2 Experiment. Its aim is to precisely measure the anomalous magnetic moment of the muon. In full operation, in order to avoid contamination, the newly born secondary beam is injected into a 505 m long Delivery Ring (DR) wherein it makes several revolutions before being sent to the experiment. Part of the commissioning scenario will execute a running mode wherein the passage from the DR will be skipped. With the aid of numerical simulations, we provide estimates of the expected performance.

## INTRODUCTION

The Muon g-2 Experiment, at Fermilab [1], will measure the muon anomalous magnetic moment, $\alpha_\mu$ to unprecedented precision: 0.14 parts per million. To perform the experiment, a polarized beam of positive muons is injected into a storage ring with a vertical uniform magnetic field. Since the positron direction from the weak muon decay is correlated with the spin of the muon, the precession frequency is measured by counting the rate of positrons above an energy threshold versus time. The g-2 value is then proportional to the precession frequency divided by the magnetic field of the storage ring.

It is important to emphasize that the delivery of a clean muon beam that has a pion contaminant fraction below $10^{-5}$, with no protons present, is a key requirement as these hadrons could cause a hadronic "flash" at injection. For this reason and after the secondary particles are produced, the beam is injected into a 505 m long Delivery Ring (DR) [2] wherein it makes several revolutions before being sent to the experiment. This will provide enough time for all pions to decay into muons and will increase the gap between the "light" muons and "heavy" protons. A kicker will then be used to remove the protons, and the muon beam will be extracted into the M4 line, and finally into the M5 beamline which terminates just upstream of the entrance of the Muon g-2 Experiment storage ring [see Fig. 1(a)]. Part of the commissioning scenario will execute a running mode wherein the passage from the DR will be skipped [see Fig. 1(b)]. With the aid of numerical simulations, we provide estimates of the expected performance of this scenario. We also outline similarities and differences between the two alternative scenarios.

## MUON CAMPUS OPERATION

Protons with 8 GeV kinetic energy are transported via the M1 beamline to an Inconel target at AP0. Within a 1.4 s cycle length, 16 pulses with $10^{12}$ protons and 120 ns full length, are arriving at the target. Secondary beam from the target will be collected using a lithium lens, and positively-charged particles with a momentum of 3.1 GeV/c (± 10%) will be selected using a bending magnet. Secondary beam leaving the Target Station will travel through the M2 and M3 lines which are designed to capture as many muons with momentum 3.094 GeV/c from pion decay as possible. The beam will then be injected into DR and the muon beam will be extracted into the new M4 line, and finally into the new M5 beamline which leads to the storage ring of the Muon g-2 Experiment. In our study the simulation terminates just upstream of entrance to that ring (end of M5).

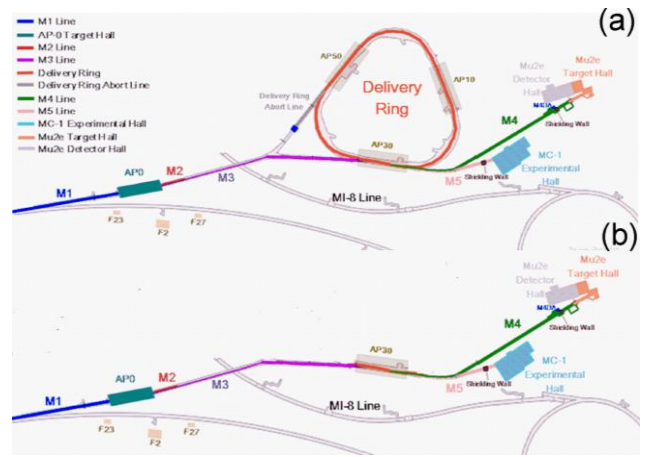

Figure 1: Layout of the Fermilab Muon Campus showing the beamlines for the Muon g-2 Experiment and Mu2e Experiment: (a) During full operation, and (b) initial commissioning plan. In that case, revolutions of the beam around the DR will be omitted.

For the full operation scenario [see Fig. 1(a)], detailed numerical simulations with two independent simulation codes [3, 4] revealed that at the end of M5, the number of

___

muons per proton on target (POT) within the ring acceptance Δp/p=±0.5% is ≈ 2.0x10⁻⁷. The beam is centered at 3.091 GeV/c with a spread Δp/p=1.2% and has an averaged polarization equal to 96%. Both codes show that the number of pions reaching the storage ring is practically zero. In the next section, we will describe in more detail the performances when revolutions around the DR are skipped.

## BEAM PERFORMANCE

The performance of the Muon Campus beamlines was simulated using G4Beamline [5]. The code includes pion decay, muon decay and tracking of the muon spin. The evolution of the number of secondary particles as function of distance is shown in Fig. 2. The plot also displays the performance variations for alternative momentum acceptances from the central value of 3.1 GeV/c. Note that $S = 0$ is just downstream of the lithium lens of the production target.

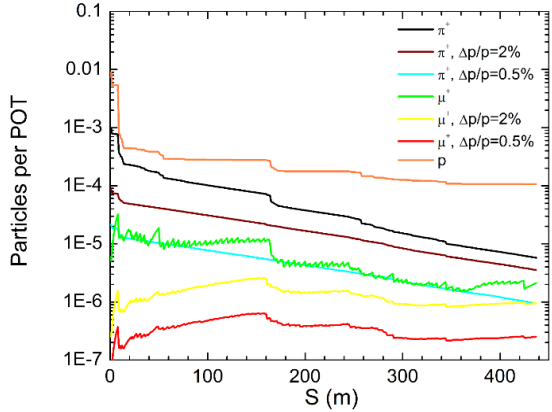

Figure 2: Simulated performance of the transmission of secondary particles. Plots for different acceptances are shown. The distance between the production target and the entrance of the storage of the Muon g-2 Experiment is 450 m approximately.

A close observation of Fig. 2 reveals a rapid loss of secondary particles within the first 10 m of the channel. This fact is not surprising since the particles produced at the target have a very wide momentum spread which extends up to 7 GeV/c. On the other hand, a selection magnet placed just downstream the target selects particles only near 3.1 GeV/c. As a result, a significant number of particles are lost. Figure 2 also indicates a significant beam loss at $S = 160.0$ m which is also the location of one of the two $9.25^0$ horizontal bending magnets in the M3 line. This loss is more evident for muons rather than pions and is independent of the momentum spread [3]. Unlike pions, the transverse phase-space of all created muons at S=160 m (calculated from second order moments) is three times larger than the beamline acceptance of 40 mm.mrad. Thus, it is very likely that the muon loss at S=160 m is due to collimation from the bending magnet at that location.

Table 1 summarizes the number of secondaries at the entrance of the storage ring of the Muon g-2 Experiment. Based on the results of Table 1 and Figure 2 the following points are noteworthy. First, the beam is dominated by secondary protons which surpass pions and muons by at least one order of magnitude. In particular, the ratio between muons and protons is 1 over 60. Second, the number of pions and muons is nearly equal at least in order of magnitude. This suggests that the beamline is not long enough to ensure a full decay of pions.

Table 1: Population of secondary particles at the end of M5 for different momentum acceptances.

| Quantity | Particle | Value (per POT) |
|---|---|---|
| Total | proton | 1.0x10⁻⁴ |
| $\Delta p/p = \pm 2\%$ | proton | 6.5x10⁻⁵ |
| $\Delta p/p = \pm 0.5\%$ | proton | 1.8x10⁻⁵ |
| Total | pion | 5.7x10⁻⁶ |
| $\Delta p/p = \pm 2\%$ | pion | 3.5x10⁻⁶ |
| $\Delta p/p = \pm 0.5\%$ | pion | 9.4x10⁻⁷ |
| Total | muon | 2.1x10⁻⁶ |
| $\Delta p/p = \pm 2\%$ | muon | 9.5x10⁻⁷ |
| $\Delta p/p = \pm 0.5\%$ | muon | 2.5x10⁻⁷ |

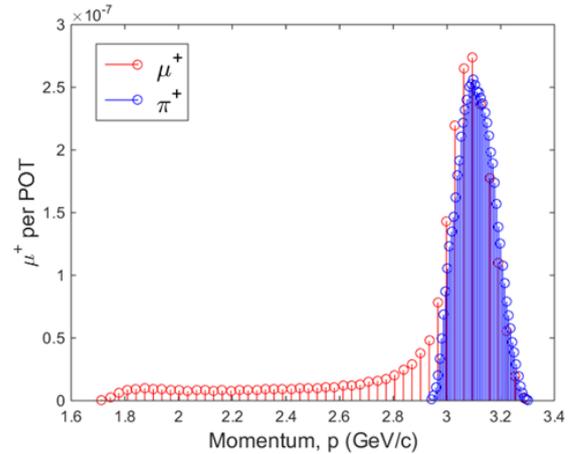

Figure 3: Histogram of the momentum distribution at the end of M5 for pions (blue) and muons (red).

Figure 3 shows a histogram of the muon and pion populations at the end of the M5 line. For simplicity, the population of protons is omitted. While the pions have an average momentum of 3.1 GeV/c with one standard deviation of 2.1 % the distribution of muons is scattered in a wide range of momenta. This is not surprising and is a direct consequence of the fact that pions are still decaying to muons and the newly born muons can have a wide range of energies.

Figure 4 shows the distribution in time for both muons and protons at the end of the M5 beamline. The top figure has the population of protons while the bottom has the population of muons. Note that both plots share the same horizontal axis. Based on Fig. 4 we can derive the following conclusions: First, both protons and muons retain the length distribution of the incoming proton beam from the Recycler [6]. Second and unlike the full operation scenario

[3], there is an overlap between the two species. This suggests that muons will have a contamination of protons as they enter the storage ring of g-2 Experiment.

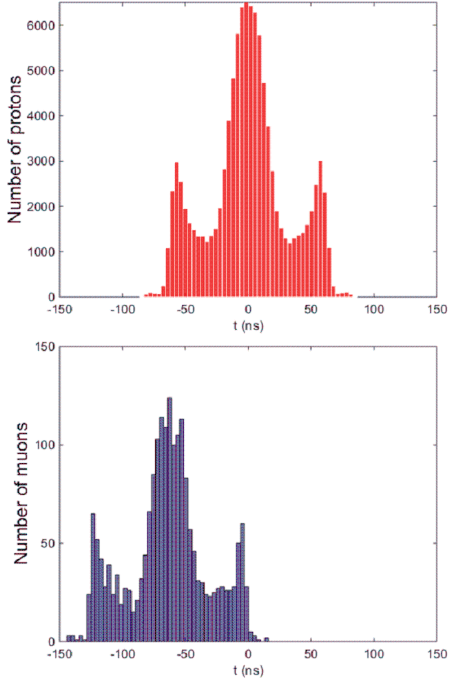

Figure 4: Longitudinal time profile of the distribution of protons (top) and muons (bottom). Both plots share the same horizontal time axis.

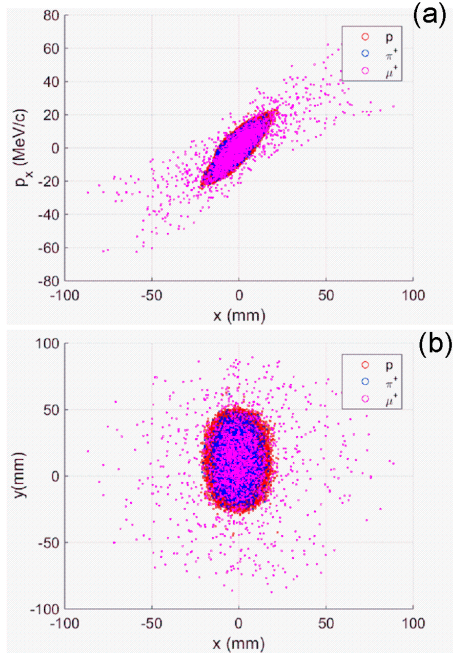

Figure 5: Secondary beam distribution at the end of M5: (a) Horizontal phase-space, and (b) configuration space.

Figure 5 shows the distribution of protons, muons and pions in phase space [Fig. 5(a)] and configuration space [Fig. 5(b)] at the end of the beamline M5. As expected the protons and pions are providing a more compact phase-space compared to muons which are scattered at a wide range of different energies. As previously described, this is a direct result of the still occurring pion decays.

Figure 6 shows the polarization of muon beam as it enters the storage ring of the Muon g-2 Experiment based on our simulation model. A polarized muon is obtained through the weak decay of pions in flight $\pi^+ \rightarrow \mu^+ + \nu_\mu$. As previously noted, the newly born muons will have a very wide spectrum of momentum. On the other hand, since the channel has a narrow momentum acceptance, i.e. $\Delta p/p = \pm 2\%$, only muons from forward decays are eventually selected. Accordingly, when the beam reaches the end of M5 it's average polarization is 96%. However, it is solely in the longitudinal direction (see Fig. 6). This last fact is very different compared to the full operation mode wherein the polarization is shared in both horizontal and longitudinal directions. This effect was found [3] to occur due to spin precession as the muon beam loops the Delivery Ring.

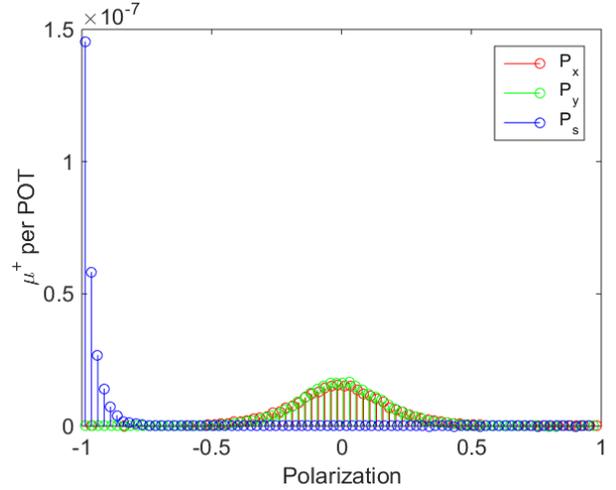

Figure 6: Simulated polarization of the g-2 muon beam as it reaches the entrance of the storage ring of the Muon g-2 Experiment. Vertical, horizontal and longitudinal components are shown. The beam is 96 polarized with a significant component only at the longitudinal direction.

## CONCLUSION

The proposed operating mode avoids delays from commissioning and operation of the DR. Thus, it has the potential to offer a quick proof-of-principle test before the Fermilab summer 2017 shutdown. Both muons and protons will be stored in the ring, but protons don't decay. Muons will decay to positrons which will be measured at the calorimeters. The overall goal is to measure the anomalous magnetic moment in a similar fashion to that described in Ref. 7. The authors are grateful to J. Annala, D. Still, B. Drendel, V. Tishchenko, and S. Werkema for many fruitful discussions.